# An analytic approach for understanding mechanisms driving breakthrough infections

Authors: Amanda Brucker[1], Jillian H Hurst[2], Emily C O'Brien[3], Deverick Anderson[4], Michael E Yarrington[4], Jay Krishnan[4], Benjamin A Goldstein[1]

1. Department of Biostatistics & Bioinformatics, Duke University, Durham, NC
2. Department of Pediatrics, Duke University, Durham, NC
3. Department of Population Health, Duke University, Durham, NC
4. Department of Medicine, Duke University, Durham, NC

Corresponding author: Ben Goldstein

## Abstract

Real world data is an increasingly utilized resource for post-market monitoring of vaccines and provides insight into real world effectiveness. However, outside of the setting of a clinical trial, heterogeneous mechanisms may drive observed breakthrough infection rates among vaccinated individuals; for instance, waning vaccine-induced immunity as time passes and the emergence of a new strain against which the vaccine has reduced protection. Analyses of infection incidence rates are typically predicated on a presumed mechanism in their choice of an "analytic time zero" after which infection rates are modeled. In this work, we propose an explicit test for driving mechanism situated in a standard Cox proportional hazards framework. We explore the test's performance in simulation studies and in an illustrative application to real world data. We additionally introduce subgroup differences in infection incidence and evaluate the impact of time zero misspecification on bias and coverage of model estimates. In this study we observe strong power and controlled type I error of the test to detect the correct infection-driving mechanism under various settings. Similar to previous studies, we find mitigated bias and greater coverage of estimates when the analytic time zero is correctly specified or accounted for.

## 1. Introduction

Vaccine effectiveness studies increasingly leverage real world data (RWD) to assess how a vaccine performs in real world settings. Analyzing vaccine effectiveness using RWD has offered insight into differential vaccine-induced immunity across diverse populations and subgroups as well as under varying schedules or series completion[1,2]. Moreover, longitudinal data captured over a longer time frame may reveal rarer outcomes or interactions with time-varying covariates[3]. However, there are challenges in studying vaccine effectiveness using RWD. The precise timing of vaccine receipt and current vaccine status of a given individual may be incomplete in electronic health records or claims databases, which only reflect the individual's healthcare encounters through particular clinics or insurers, introducing the risk for misclassification and resulting in bias[4,5]. Data elements such as medications, laboratory measurements, vital signs, and comorbidities may be documented at a more regular cadence among specific patient subgroups (e.g., sicker patients, patients with more comprehensive insurance coverage) so that their presence is not only informative but may act as a collider or confounder in the association of interest[6]. Furthermore, as in all observational studies, unobserved or observed patient characteristics may confound the association between vaccine and outcomes. Many prior vaccine effectiveness studies that utilize RWD have addressed the complexities above by focusing on a vaccinated-only population[7], utilizing data sources with consistent capture of the characteristics of interest[8], and employing standard regression methods to adjust for potential confounders[9].

In effectiveness studies of vaccinated-only populations, the focus is often on characterizing the incidence and relative incidence of breakthrough infections as in Goldberg et al.[1] and Coburn et al.[10]. In such studies, it is implicit that infections may be driven by different mechanisms, e.g., naturally declining immunity following vaccination or a dominant new strain. This uncertainty is accounted for by focusing on a narrow time window to hold calendar-time-varying mechanisms constant or by constructing patient cohorts based on the timing of vaccination. However, little work has been done to develop methods that explicitly test for the mechanism driving breakthrough infections or explore the performance of standard approaches under a variety of infection-generating signals.

In this work, we propose a novel analytic framework for studying vaccine effectiveness using RWD. Our work is motivated by efforts to assess the effectiveness of COVID-19 vaccine during different waves of the pandemic. However, the analytic framework derived is appropriate for any pathogen that exhibits cyclical variation in the presence of vaccinated populations (e.g., influenza virus). We leverage a standard time-to-event framework and introduce a vaccination offset variable to accommodate possible misspecification of time-zero. This framework allows for simultaneous inference on the driving mechanism behind breakthrough events as well as potential subgroup differences in event rates. We evaluate the framework in a simulation study under a variety of settings and illustrate use to evaluate COVID-19 vaccine waning among patients from our health system.

## 2. Methods

**2.1 Motivation**

We consider a vaccinated population in which we observe breakthrough infections and assess two questions: 1) what mechanism is causing the breakthrough infections; and 2) accounting for the mechanism, are there differential rates of breakthrough infections in population subgroups? In addressing these questions, we posit that breakthrough infections can arise via one of two mechanisms: 1) the vaccine effectiveness has waned due to naturally declining immunity following vaccination or 2) the vaccine is no longer effective against the strain of virus currently in circulation.

As we develop below, a natural framework to address these questions is a time-to-event framework. A time-to-event analysis requires the specification of a time origin or time zero, which is traditionally chosen to represent the time at which an individual is first at risk for the outcome. In studies of breakthrough infections in a vaccinated population, the date of vaccination is a common choice for time zero since individuals become at risk for the event no earlier than this date. However, if the hypothesized driving mechanism for an observed time period of breakthrough infections is a new viral strain, a time-to-event analysis should be indexed with respect to a calendar date corresponding to the emergence of the strain. As illustrated in Figure 1, incorrectly defining time zero will lead to inaccurate estimates of incidence and relative incidence. As such, one of the analytic challenges in this work is to properly index a time-to-event model to account for an *unknown* breakthrough mechanism.

**Figure 1. Kaplan-Meier curves for breakthrough infections under two choices of time zero.** We show Kaplan-Meier survival estimates for 2 subgroups in a simulated dataset. The driving mechanism is a new strain. We obtain survival curves estimates specifying time zero as individuals' vaccination dates or a calendar date corresponding to the emergence of the new strain.

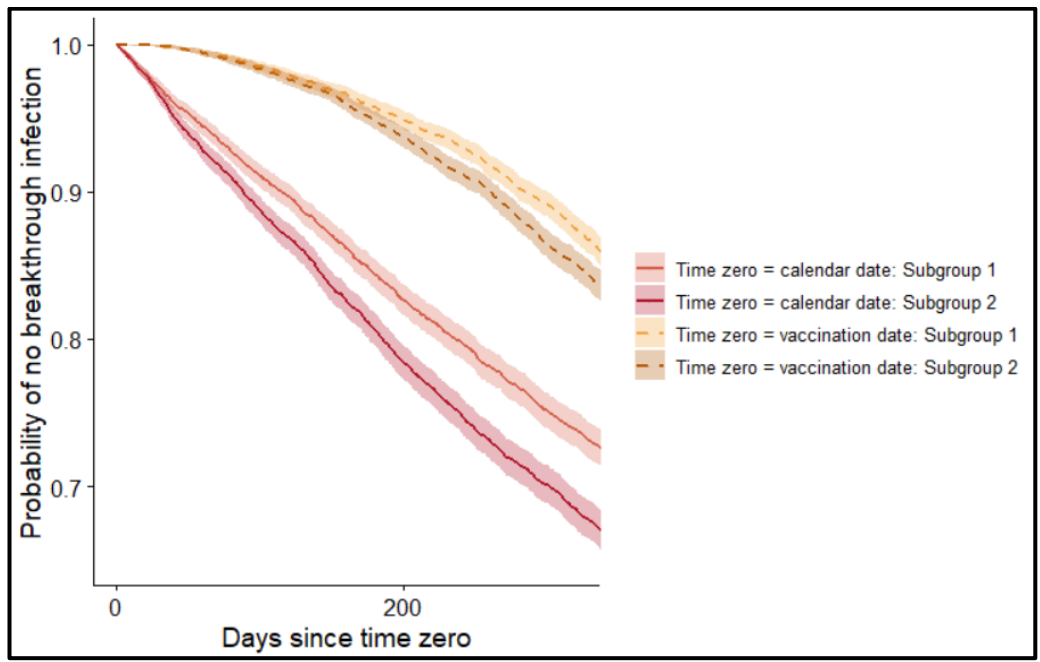

## 2.2 Analytic Framework

Under the two possible mechanisms above, we consider a time-to-event based analysis. We assume patient-level data, including date of vaccination and associated characteristics (e.g., demographic, socioeconomic, baseline clinical) $x_i = (x_{i1}, x_{i2}, ..., x_{ip})$. We take the characteristics to be time invariant but the framework can be modified to accommodate time-varying features as well. We designate a landmark date L after which we are interested in assessing the risk of breakthrough infection, and we specify an administrative censoring date after which we are no longer interested in assessing risk. We additionally define a vaccination period start date and identify the set of people who are vaccinated between this date and the landmark date, and who have remained event free up through time L. For these individuals we generate an event indicator $C_i$ where $C_i$ = 1 indicates an observed breakthrough infection and $C_i$ = 0 indicates censoring. We define $T_i > 0$ as the time from the landmark date to the patient's event or censor date. We further define the variable $z_{i\Delta} > 0$ as the days between the patient's vaccination and the landmark date, as illustrated in Figure 2.

**Figure 2. Illustration of proposed framework.**

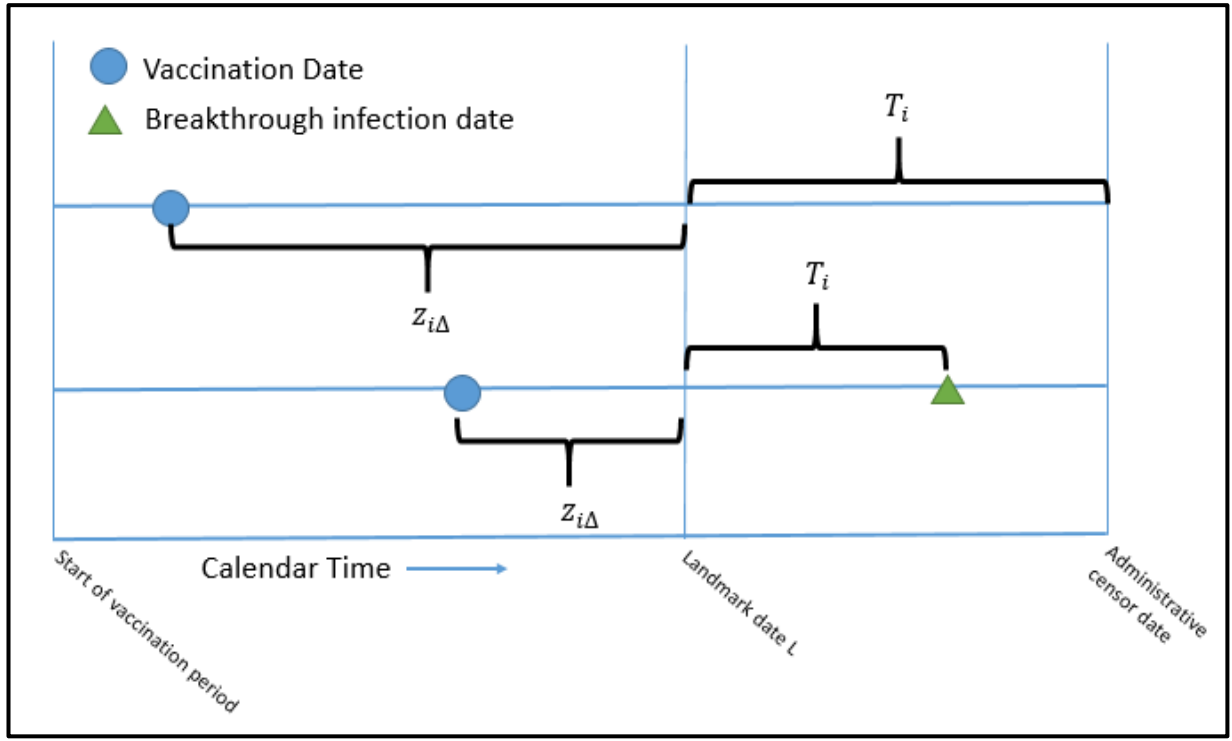

We perform Cox proportional hazards (PH) regression based on hazard function shown in Equation 1 which corresponds to a standard Cox PH regression with $x_i$ and $z_{i\Delta}$ as independent variables. In this framework, our parameter of interest is $\beta_\Delta$, which reflects the association between vaccination recency and hazard of infection. We test the null hypothesis shown in Equation 1: i.e., the hypothesis that there is no association between recency of vaccination and hazard of infection.

$$h(t|x_i, z_{i\Delta}) = h_0(t) \times \exp(\beta_\Delta z_{i\Delta} + \beta_1 x_{i1} + \beta_2 x_{i2} + \cdots \beta_p x_{ip}) \quad \text{Equation 1}$$

$$H_0: \beta_\Delta = 0 \text{ vs } H_A: \beta_\Delta \neq 0$$

Under the assumption that there are only 2 possible breakthrough mechanisms, the rejection of this null suggests that waning immunity is contributing to the infections observed following the landmark date L. We note that the failure to reject the null does not confirm that breakthrough infections are due to a new strain or not due to waning immunity. In the framework above, we can simultaneously obtain estimates of $\beta_1, \beta_2, \cdots, \beta_p$, which allows for standard inference on subgroup differences and continuous effects of covariates.

**2.3 Simulation study**

We conducted a simulation study to 1) assess the power and type one error rates of the hypothesis test in Equation 1, providing inference on our first question of interest; and 2) examine the bias of subgroup differences estimates $\beta_1, \beta_2, \cdots, \beta_p$ obtained from the analytic framework above, corresponding to our second question of interest.

*2.3.1 Assessing Underlying Breakthrough Mechanism*

We simulated data under two proposed breakthrough mechanisms. The first mechanism, $M_W$, generated breakthrough events due to naturally declining immunity following vaccination, and corresponded to a desire to *reject* the null hypothesis $\beta_\Delta = 0$. Under $M_W$, we assumed the following:

- Immediately following vaccination, the individual had the lowest daily risk of infection **a**;
- The daily risk was constant at its lowest level for duration **d** days following vaccination;
- The daily risk then grew linearly at rate **r** until the individual's daily risk returned to pre-vaccination levels **b**.

The second mechanism, $M_S$, was the emergence of a new strain that heightened an individual's risk of infection and uncoupled risk from the date of vaccination. When infections were generated under $M_S$, we expected to *fail* to reject the null hypothesis. Under $M_S$, we specified:

- The patient's daily risk of infection was at a constant level **k** prior to the new strain
- At the emergence of the new strain, the risk jumped to a constant level **c** greater than the pre-strain risk.

Each of these mechanisms corresponded to a different hazard function represented in **Figure S1**.

To mimic real world data, the simulations were anchored around an arbitrary window of calendar time. We generated uniformly random vaccination dates across a 365-day period and set the landmark date L as the end of the period. We applied an administrative censoring date 365 days following the landmark date L. Individuals without an infection during this 365-day period were censored.

*2.3.2 Assessing Subgroup Differences*

We additionally evaluated the estimation of subgroup differences under both mechanisms. Using the notation in Equation 1, we simulated a single binary variable $x_1$ indicating subgroup membership with possible values 0 or 1. No other categorical or continuous variables were generated, i.e., $x_2, \ldots, x_p = NULL$. We simulated data in which the hazard was consistent across the two subgroups ($\beta_1 = 0$) and data in with different hazards between the subgroups ($\beta_1 \neq 0$). We applied an unadjusted Cox PH regression to the former setting, with $z_\Delta$ as the only independent variable, and an adjusted Cox PH regression to the latter setting with $z_\Delta$ and the subgroup indicator $x_1$ as independent variables. In data with subgroup differences, we randomly assigned 50% of all patients to a subgroup (captured in a binary indicator $x_1$) and let this subgroup modify patients' hazard functions through a proportional term $\exp(\beta_1 x_1)$. Thus, $\beta_1$ translated to the log hazard ratio between the two subgroups. We assessed the bias and coverage of the model-estimated $\widehat{\beta_1}$ with respect to the true $\beta_1$. We additionally assessed these measures from two naïve Cox PH regressions with only $x_1$ as an independent variable, one designating time zero as the milestone date and one designating time zero as the vaccination date.

*2.3.3 Simulation Settings*

We simulated B = 1000 analytic datasets under different parameter settings corresponding to varying event (hazard) rates, immunity waning rates, sample sizes, and presence of subgroup differences (**Table 1**). Across all datasets, an overly large sample size was simulated and individuals who had an event prior to the landmark date were left-truncated as would occur in a real data analysis. A random sample from the remaining individuals was taken obtain the desired sample size. In datasets with subgroup differences, we set $\beta_1$ = 0.15. We calculated the power of the hypothesis test as (# p-values ≤ α)/B when the null hypothesis should be rejected (i.e., data simulated under $M_W$). We calculated the type I error of the hypothesis test as (# p-values ≤ α)/B for when the null hypothesis should not be rejected (i.e., data simulated under $M_S$).

**Table 1. Simulation settings.**

|  | No subgroup differences | With subgroup differences |
|---|---|---|
| ***Waning immunity ($M_W$)*** |  |  |
| Lowest hazard | a = 0.0001 | a = 0.0001 |
| Highest hazard | b = 0.0007 | b = 0.0007 |
| Duration at lowest hazard | d = (90, 180, 240 days) | d = (90, 180, 240 days) |
| Rate of declining immunity (increase in hazard per day) | r = (1e-06, 5e-06, 1e-05, 5e-05, 1e-04) | r = (1e-08, 1e-07, 1e-06, 1e-05) |
| ***New strain ($M_S$)*** |  |  |
| Flat hazard | c = (1e-04, 5e-04, 1e-03, 5e-03, 1e-02) | c = (1e-04, 5e-04, 1e-03, 5e-03, 1e-02) |
| ***Method evaluation at each combination of the above settings*** |  |  |
| Analytic dataset sample size | N = (500, 1000, 10000, 100000) | N = (500, 1000, 10000, 100000) |
| Significance level for power and type I error | α = (0.01, 0.05, 0.10) | α = (0.01, 0.05, 0.10) |

## 2.4 Real world data illustration

To illustrate the analytic approach, we applied the proposed above framework to a real-world data set and evaluated breakthrough COVID-19 infections among vaccinated individuals in our health system. We identified individuals who were vaccinated between 1/1/21 (the beginning of vaccine administration) and 7/1/21 (prior to the emergence of the SARS-CoV-2 Delta variant). Vaccine information was collected through linkage of our health system's EHR data with the state vaccine reporting system. The outcome of interest was a positive COVID-19 test result recorded in the EHR. We limited the cohort to individuals living in the county in which our health system is located and individuals with at least one encounter in our health system in the prior two years, to increase the likelihood of observing their test results in the EHR. We abstracted outcomes in the time window from 7/1/21 – 12/1/21, corresponding to a period when the Delta variant was dominant and prior to the emergence of the Omicron variant. We extracted demographics (age, sex, race-ethnicity) and chronic health conditions (chronic obstructive pulmonary disease [COPD], renal disease, liver disease, hypertension, diabetes, asthma, cardiovascular disease [CVD], cancer) for each individual in the cohort (see **Table S2** for ICD-10-based phenotyping of health conditions). We tested for the mechanism driving breakthrough infections and estimated subgroup differences across patient demographics and chronic diseases. We report the results of the test for driving mechanism as well as hazard ratios (HR) and 95% confidence intervals (95% CI) for subgroup differences.

All analyses were performed in R version 4.0.2. The simulation code is available in the Supplement.

## 3. Results

### 3.1 Illustration of vaccination offset variable and breakthrough infections under different mechanisms

We first explored breakthrough infections generated under the two simulation mechanisms in a Kaplan-Meier survival analysis (**Figure 3**). Infections were simulated under the condition of no subgroup heterogeneity and time-to-infection was defined with respect to the calendar milestone date as described in the Methods section. Patients were grouped into cohorts based on their $z_\Delta$ variable, in 30-day bins spanning patients vaccinated within 30 days prior to the milestone date to patients vaccinated > 180 days prior. In panel A, infections driven by naturally declining immunity show a fanning effect in which patients with the smallest $z_\Delta$ (vaccinated most recently) have the lowest infection rate (light red curve) and the cohort of patients with the largest $z_\Delta$ (longest time since vaccination) have the highest infection rate (dark red curve). In panel B, the incidence of breakthrough infections driven by a new strain does not appear to vary across cohorts. Thus, we observe a gradient of survival curves as $z_\Delta$ varies when declining immunity is driving infections, and no gradient when a new strain resets the hazard of infection.

**Figure 3. Kaplan-Meier survival curves for simulated breakthrough infection data under two different mechanisms.** This figure illustrates survival curve estimates for cohorts of patients stratified by $z_\Delta$, i.e., their recency of vaccination at the milestone date. The different colored curves correspond to cohorts of patients based on their recency of vaccination. A) The true mechanism driving these simulated breakthrough infections is naturally declining immunity following vaccination. B) The true mechanism driving these simulated breakthrough infections is a new strain emerging.

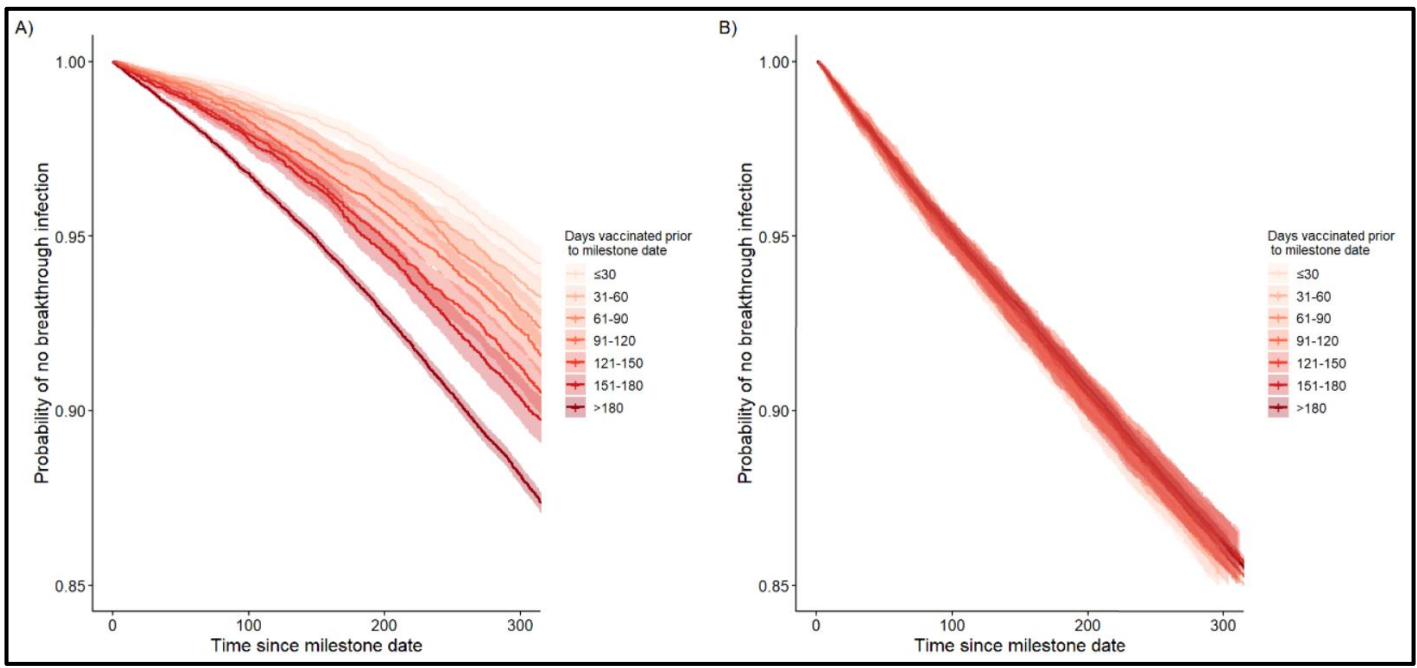

## 3.2 Simulation study results

*3.2.1 Test for driving mechanism*

The type I error rate of the proposed test is shown in **Table S1**. The Type I error was controlled at the nominal levels under all settings with no apparent trends across varying simulation parameters. Error was controlled under simulated data with and without subgroup heterogeneity. The power of the test is visualized in **Figure 4,** under no subgroup heterogeneity and an alpha of 0.05. We observed that the test achieves perfect or near perfect power under all settings when the sample size is 10,000 or larger. Power was slightly greater under a shorter duration of maximum immunity. We observed similar trends in power curves corresponding to data with subgroup heterogeneity (**Figure S2**).

**Figure 4. Power of proposed test.** Power is calculated based on data simulated under naturally declining immunity. Power is shown on the y-axis and the waning immunity rate varies along the x-axis. Power was evaluated at 5 values of waning immunity rate; 3 different durations of maximum immunity shown in red, green, and blue curves; and 4 different sample sizes (4 sub-panels). These curves correspond to power under alpha=0.05 and no subgroup heterogeneity.

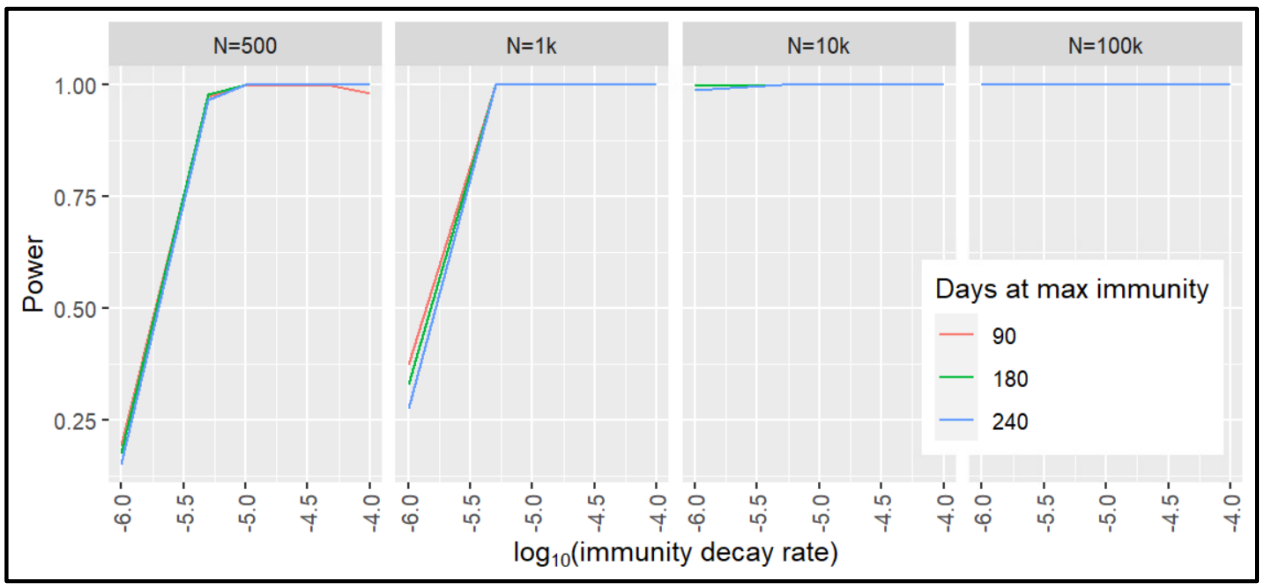

*3.2.2 Estimation of subgroup differences*

We explored the bias of subgroup estimates under a signal arising from a new strain, shown in **Figure S3**. The proposed method produced nearly identical estimates to those from a naïve Cox PH regression, suggesting that the inclusion of the vaccination offset variable does not significantly impact the performance of a standard survival analysis when the time zero is correctly specified. We calculated the bias of subgroup estimates under waning immunity along with their coverage as shown in **Figure 5**. Under the slowest immunity decay rate settings, the proposed method and calendar time Cox PH regression methods performed similarly. As immunity waning rates increased, the proposed method produced less biased estimates than the naïve approach. Across all settings, the vaccination time Cox PH regression produced biased estimates. The coverage under the proposed method and the vaccination time Cox PH regression consistently hovered around 95%, while the coverage under the calendar time Cox PH regression decreased sharply when immunity decay rates were higher.

**Figure 5. Bias and coverage of subgroup difference estimates under waning immunity.** Panel A) Mean bias is shown on the y-axis and the waning immunity rate varies along the x-axis. B) The percent of estimates whose 95% confidence interval covered the true value (coverage) is shown on the y-axis and the waning immunity rate varies along the x-axis. Both metrics were calculated at 5 values of waning immunity rate; 4 different sample sizes (4 sub-panels); and a duration of maximum immunity of 180 days.

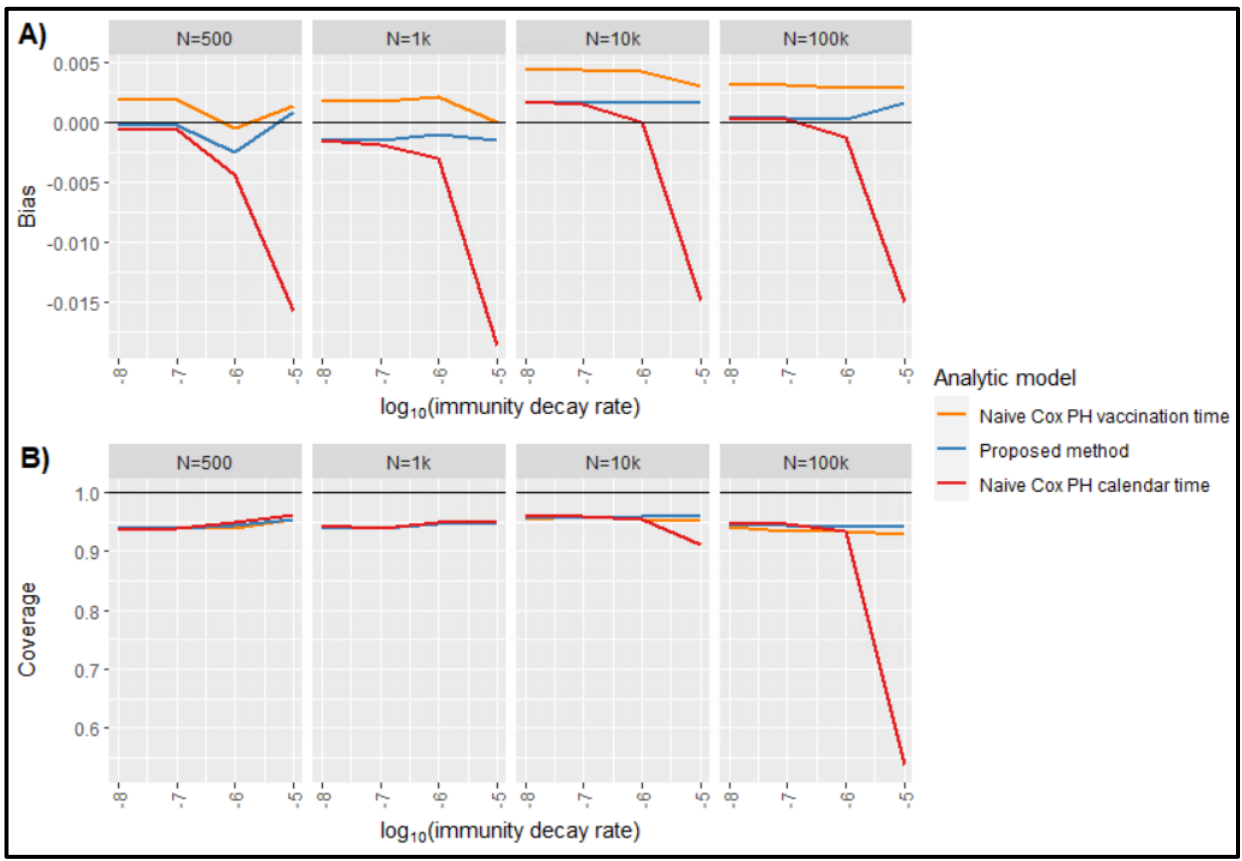

## 3.3 Real world data illustration results

The real-world data analysis included 96,158 vaccinated individuals with 509 breakthrough infections between 7/1/21 – 12/1/21. We visualized the incidence of breakthrough infections in **Figure 6,** stratified by the recency of vaccination or 30-day bins of $z_\Delta$. We constructed 6 cohorts of patients who were vaccinated within 1-month increments of calendar time ranging from January 2021 (maroon) to June 2021 (pale red). The significant overlap of these survival curves along with some fanning behavior exhibits similarities with the survival curves observed under both of the simulation study mechanisms (**Figure 3**).

**Figure 6. Kaplan-Meier curves of breakthrough infections in real world data.** The breakthrough infection rates are illustrated in 6 cohorts of patients constructed based on the recency of their vaccination with respect to the milestone date of 7/1/21. This is equivalent to stratifying by the vaccination offset variable $z_\Delta$,

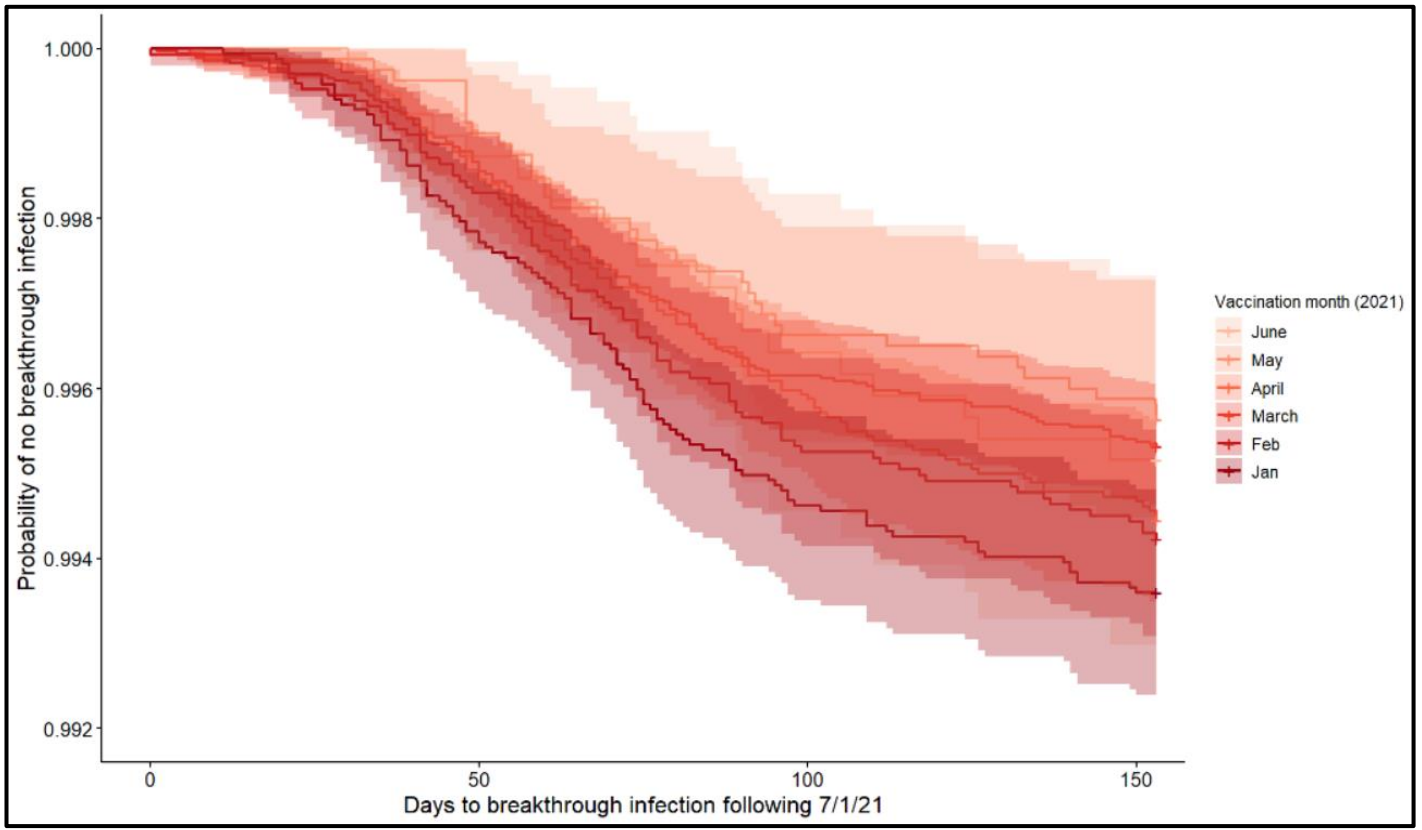

We applied the proposed method to test whether waning immunity may be contributing to infections observed after emergence of the Delta variant. As shown in **Table 2**, the test was marginally significant with a p-value = 0.040 and the estimated HR (95% CI) was 1.003 (1.000 – 1.005), suggesting that breakthrough infections in 7/1/21-12/1/21 are driven partially by naturally declining immunity. We additionally performed a sensitivity test by removing patients in the earliest waves of vaccinations, keeping patients who were vaccinated within 90 days prior to 7/1/21 ($z_\Delta \leq 90$). In this population, the recency of vaccination was no longer informative.

**Table 2. Test for driving mechanism in real world data.** These results are obtained from a fully adjusted model including the vaccination offset variable as well as patient demographics and comorbidities.

| Population | # events / # patients | Vaccination offset variable Hazard Ratio* (95% CI) | P-value |
|---|---|---|---|
| All patients | 509 / 96158 | 1.003 (1.000, 1.005) | 0.040 |
| Sensitivity: Recently vaccinated patients ($z_\Delta \leq 90$) | 149 / 28646 | 0.998 (0.990, 1.005) | 0.507 |

*This corresponds to an estimate of $\exp(\beta_\Delta)$ in the model given in Equation 1.

In **Figure 7**, we visualize the HRs and 95% CIs for all covariates included in the model fit to the full population. As a comparison, we also applied a standard Cox PH regression to the data. We observe that the proposed method and Cox PH methods produce similar estimates and similar inference with the exception of Hispanic patients. In alignment with the behavior observed in the Kaplan-Meier analysis and test for driving mechanism, the subgroup difference estimates demonstrate qualities that are a mix of the subgroup estimation observed under both $M_W$ and $M_S$ in the simulation study. Some estimates obtained under the proposed method may have smaller bias that those arising from the naïve approach, although we do not know the true subgroup differences in this population. HRs, 95% CIs and p-values are tabulated in **Table S3**.

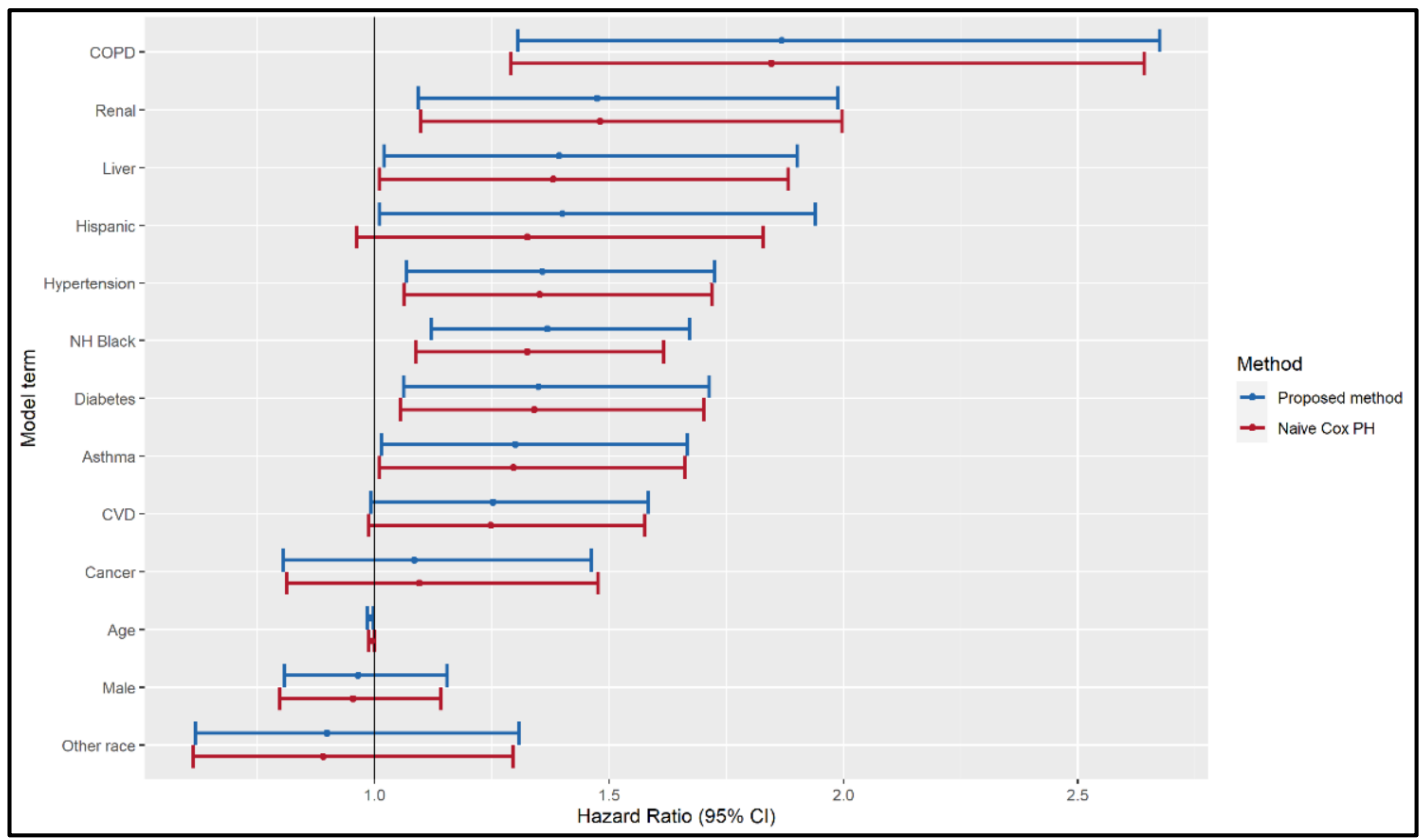

**Figure 7. Adjusted hazard ratios obtained from proposed method and naïve approach.** Adjusted hazard ratios and confidence intervals are shown for the proposed method (blue) and naïve Cox PH regression (red).

## 4. Conclusions

In this simulation study, we proposed a framework to assess and account for the underlying mechanism for a post-vaccine breakthrough infection. We framed a test for driving mechanisms as a test for the influence of waning immunity in modeling breakthrough infections with respect to calendar time. We employed a Cox PH regression framework with the addition of a vaccination offset variable to account for the potential mis-specification of time zero. Overall the approach has good power and type I error control under a variety of settings. We also observed lower bias in subgroup difference estimates under the proposed method compared to two standard Cox PH approaches when the calendar time zero was mis-specified for the mechanism. When simulated infections were driven by a new strain, the vaccination offset variable was not informative and produced estimates equivalent to a correctly specified Cox PH regression.

In an exploratory application to COVID-19 breakthrough infections in RWD, we observed a slight gradient in Kaplan-Meier survival curves based on the recency of an individual's vaccination, a mixture of what we observed under the two simulation study mechanisms. The test for driving mechanism suggested that waning immunity following vaccination had a marginal but significant effect on breakthrough infections observed after the delta variant emerged. Given this, it was not surprising that the fully adjusted subgroup HRs were similar to those obtained under naïve Cox PH regression, behavior observed in the simulation study under a dominant new strain signal.

The ability to monitor vaccine effectiveness is an important part of public health efforts to combat infectious diseases at the population level[11]. While randomized controlled trials are the gold standard for evaluating the efficacy of a vaccine in preventing infections for a given pathogen, the individuals who participate in these trials do not represent the population at large in terms of age, health conditions, and other factors that strongly influence vaccine effectiveness. Clinical trials are also unable to address changes in transmission dynamics, pathogen evolution, and human behavior that have a significant impact on infection risk[12]. In contrast, post-market monitoring using RWD can identify sub-groups that may be particularly at risk, reflect vaccine effectiveness in real world conditions (e.g., imperfect adherence to schedule or series), and allow for longer follow-up to observe breakthrough infections. Furthermore, infectious agents can evolve rapidly, as has been observed with the COVID-19 pandemic, and RWD-based monitoring of vaccine effectiveness is critical to determining when and how to modify vaccine dosing schedules and to identify the need for updated vaccine products.

The limitations of using RWD for vaccine effectiveness monitoring are also increasingly recognized. Common RWD sources like EHR, insurance claims, and administrative databases each have limited catchment and may imperfectly represent vaccine status and any breakthrough infections, leading to issues of misclassification and missing outcomes[4]. In addition, some vaccinations may be uniquely under-captured in EHR and healthcare system databases if they are frequently administered in pharmacies (e.g., influenza) or in public health departments (e.g., prenatal Tetanus, Diptheria, and Pertussis [TDAP]). The population represented in the data source might be biased to include individuals who are healthier and younger (e.g., commercial claims database), older (e.g., Medicare claims), or who have more interaction with a health system due to illness or greater engagement in managing their health. Community infection rates are also not captured in most RWD sources and individuals may be vaccinated at varying times, leading to complex interplay between time-varying risk of infection due to exposure and time-varying vaccine-conferred protection. Researchers have developed strategies to address these limitations including merging data from multiple sources, analysing populations that are most likely to be well observed in the data, assessing bias in the construction of the analysis cohort and warning about the generalizability of findings as appropriate, and anchoring the analysis to calendar time to control for temporal trends.

This work highlights the impact of a mis-specified time zero in a time-to-event framework. We examined one possible direction of mis-specification, in which the analytic time zero was set to be a calendar date when infections are driven by waning vaccine-induced immunity. We observed that a naïve approach that ignores this mis-specification results in biased estimates that worsen as the vaccine-induced immunity decays at a faster rate. Adjusting for the incorrect time zero through the proposed vaccination offset variable resulted in less biased estimates under most simulation settings. Other work has explored the topic of time zero specification and its impact on the estimation of covariate or subgroup effects. In epidemiologic cohort studies, researchers found that incorrectly specifying time zero with respect to study enrollment date rather than participant age led to biased estimates.[13,14] In the case of a mis-specified time-scale, Thiebault et al.[15] explored a similar model to ours with the appropriate time-scale included as a covariate. They found a significant reduction in bias under this model compared to a naïve unadjusted approach, as we also observed in our simulation study. However, in a subsequent study, Pencina et al.[16] found that bias from a mis-specified time-scale could be nearly eliminated through appropriate adjustment.

This study has some limitations. We evaluated the model's performance under relatively simple mechanisms and did not vary community infection rates or explicitly model viruses with different levels of transmissibility. We assumed a piecewise linear curve in the declining immunity following vaccination, when biological models might suggest an exponential decay. We did not account for patients' vaccine series in the simulations or real world data analysis, simply considering their first vaccine. A limitation of the RWD illustration is that it relied on capture of positive tests in the EHR of one healthcare system, and as a result, the absolute infection rates are lower than expected.

Overall, this work has analytic implications for how we use RWD to monitor vaccine effectiveness. By manipulating the time zero specification, we show that we can uncover the source of waning immunities and avoid biased effect estimates. Future work should consider more complicated scenarios, such as waning due to multiple mechanisms.

# Supplemental Materials

**Table S1: Type I error of proposed test.**

| # Patients | Hazard (*c*) | Type I error: test without subgroup heterogeneity | | | Type I error: test with subgroup heterogeneity | | |
|---|---|---|---|---|---|---|---|
| | | Alpha = 0.01 | Alpha = 0.05 | Alpha = 0.10 | Alpha = 0.01 | Alpha = 0.05 | Alpha = 0.10 |
| 500 | 1e-04 | 0.011 | 0.034 | 0.078 | 0.01 | 0.045 | 0.093 |
| 500 | 5e-04 | 0.006 | 0.04 | 0.088 | 0.007 | 0.048 | 0.093 |
| 500 | 1e-03 | 0.01 | 0.055 | 0.096 | 0.014 | 0.058 | 0.101 |
| 500 | 5e-03 | 0.014 | 0.05 | 0.097 | 0.01 | 0.049 | 0.103 |
| 500 | 1e-02 | 0.011 | 0.057 | 0.104 | 0.012 | 0.055 | 0.108 |
| 1000 | 1e-04 | 0.011 | 0.056 | 0.113 | 0.003 | 0.06 | 0.104 |
| 1000 | 5e-04 | 0.009 | 0.043 | 0.104 | 0.01 | 0.041 | 0.087 |
| 1000 | 1e-03 | 0.01 | 0.045 | 0.087 | 0.006 | 0.043 | 0.088 |
| 1000 | 5e-03 | 0.013 | 0.066 | 0.118 | 0.009 | 0.042 | 0.088 |
| 1000 | 1e-02 | 0.013 | 0.07 | 0.121 | 0.007 | 0.039 | 0.088 |
| 10k | 1e-04 | 0.014 | 0.047 | 0.096 | 0.007 | 0.054 | 0.103 |
| 10k | 5e-04 | 0.01 | 0.047 | 0.092 | 0.009 | 0.055 | 0.1 |
| 10k | 1e-03 | 0.002 | 0.047 | 0.1 | 0.014 | 0.054 | 0.099 |
| 10k | 5e-03 | 0.01 | 0.047 | 0.092 | 0.007 | 0.047 | 0.099 |
| 10k | 1e-02 | 0.006 | 0.041 | 0.089 | 0.008 | 0.043 | 0.098 |
| 100k | 1e-04 | 0.011 | 0.047 | 0.09 | 0.014 | 0.053 | 0.107 |
| 100k | 5e-04 | 0.011 | 0.065 | 0.119 | 0.011 | 0.051 | 0.107 |
| 100k | 1e-03 | 0.01 | 0.051 | 0.108 | 0.008 | 0.047 | 0.115 |
| 100k | 5e-03 | 0.011 | 0.054 | 0.101 | 0.007 | 0.056 | 0.105 |
| 100k | 1e-02 | 0.005 | 0.053 | 0.107 | 0.009 | 0.061 | 0.113 |

**Figure S1. Parameterized hazard curves for simulated breakthrough infections.**

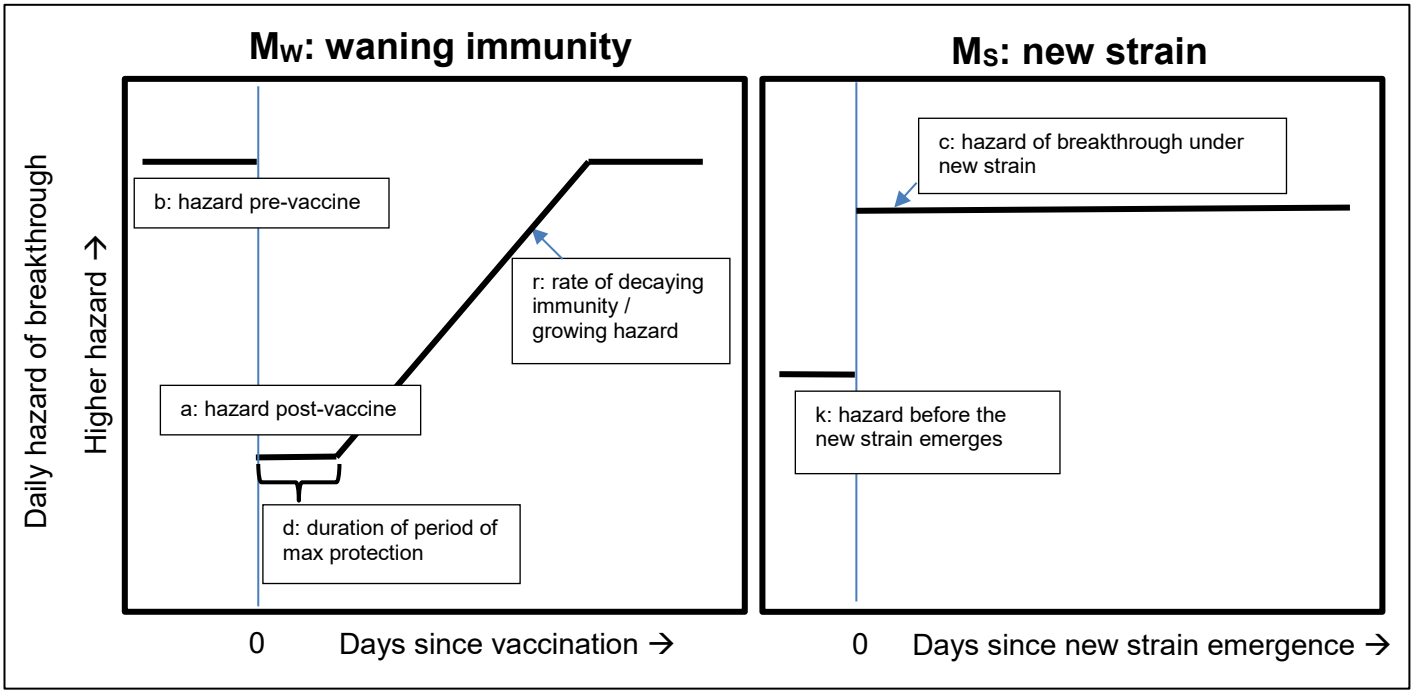

**Figure S2. Power of proposed test with subgroup heterogeneity.** Power is calculated based on data simulated under a waning immunity mechanism (MW). Power is shown on the y-axis and the waning immunity rate varies along the x-axis. Power was evaluated at 5 values of waning immunity rate (see Table 1); 3 different durations of maximum immunity shown in red, green, and blue curves; and 4 different sample sizes (4 sub-panels). These curves correspond to power under alpha=0.05.

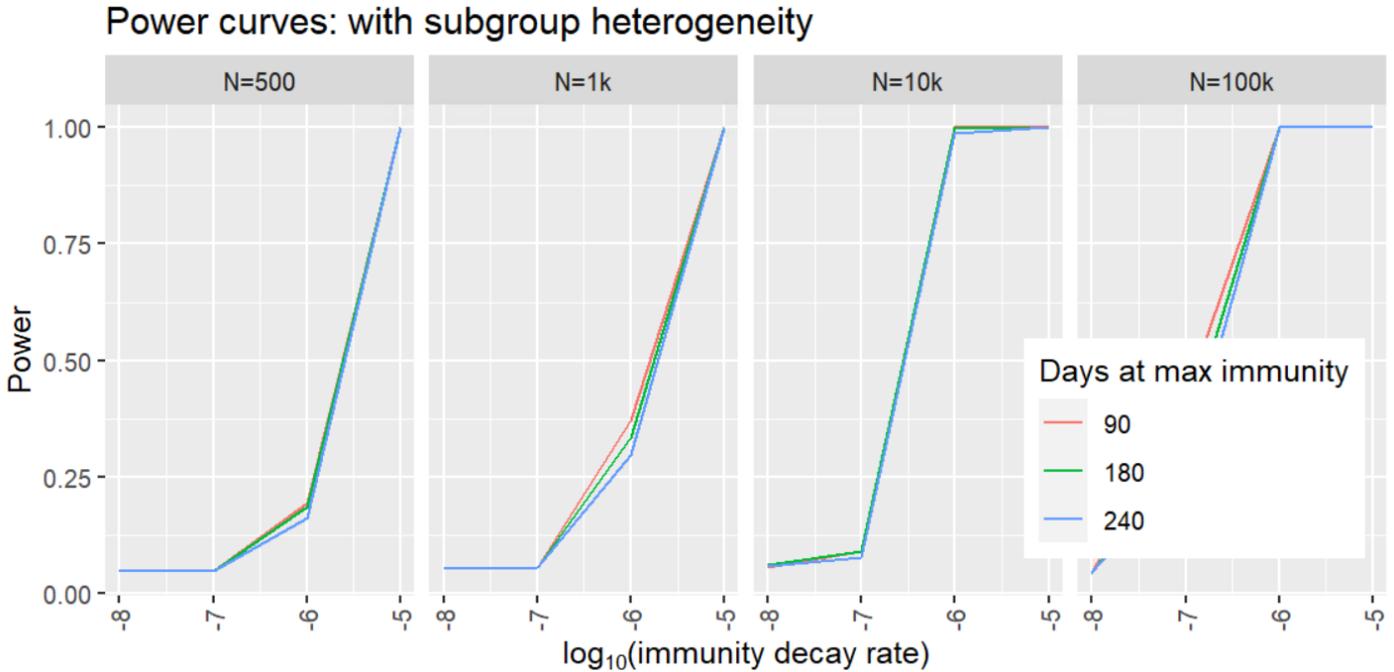

**Figure S3. Bias in subgroup difference estimates under new strain.** Mean bias across the 1000 simulation replications is shown on the y-axis and the new strain daily hazard varies along the x-axis. Bias was evaluated at 5 values of new strain hazard (see Table 1); 4 different sample sizes (4 sub-panels).

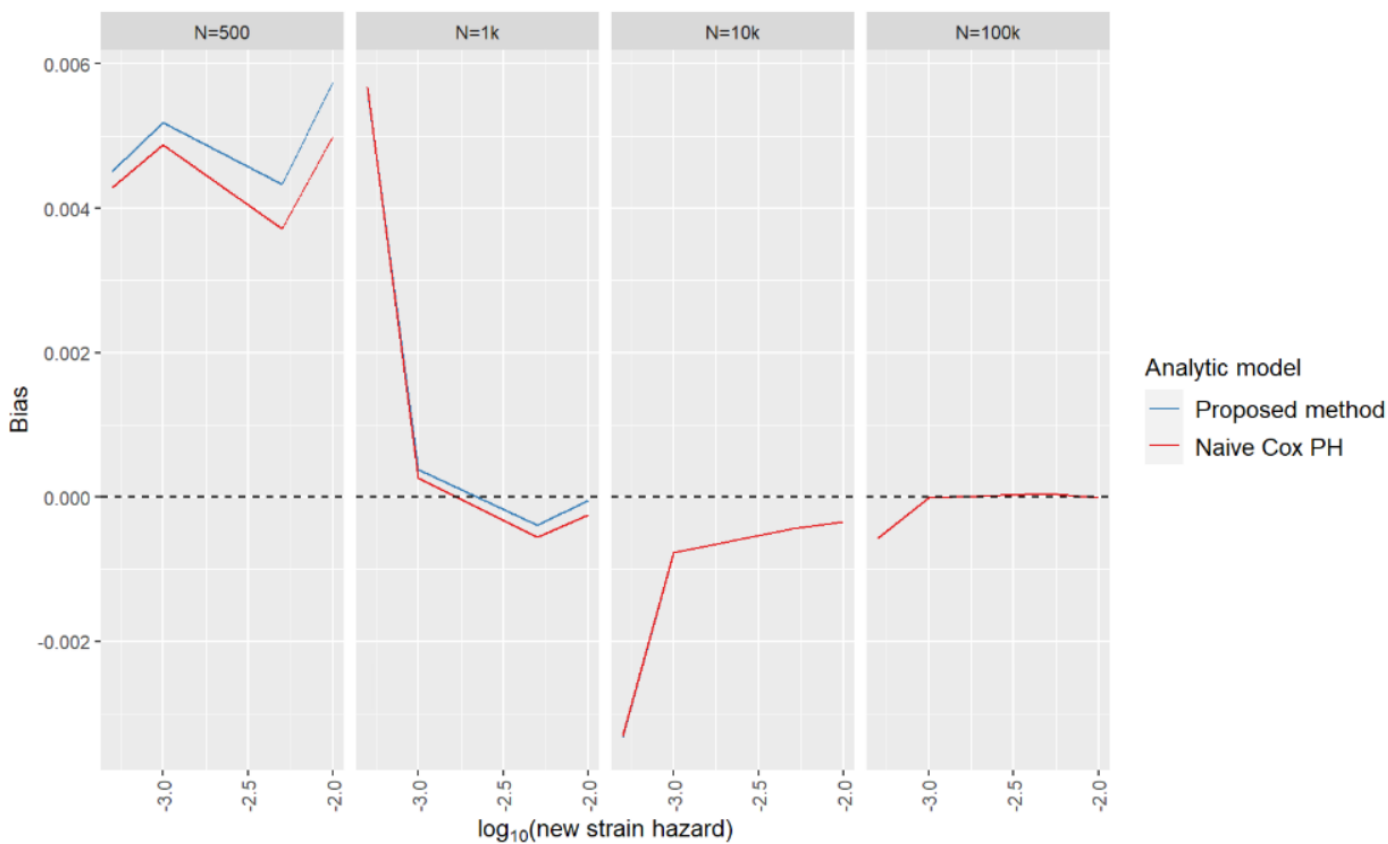

**Table S2. ICD codes to define chronic diseases in real world data application.**

| Condition | ICD-10 codes |
|---|---|
| Cancer | C%* |
| Cardiovascular disease | I[2-5][0-9]% |
| Hypertension | I1[0-1]% |
| Liver disease | K7[0-7]% |
| COPD | J44% |
| Asthma | J45% |
| Renal disease | N18% |
| Diabetes | E0[8-9]%, E1[0-3]% |

*"%" is a wildcard character that indicates any or no subsequent characters following the specified alphanumeric(s)

**Table S3. Real world data application hazard ratios and 95% confidence intervals.**

|  | Proposed method |  | Cox PH regression |  |
| --- | --- | --- | --- | --- |
| Term | HR (95% CI) | p-value | HR (95% CI) | p-value |
| Vaccination offset $z_\Delta$ | 1.003 (1, 1.005) | 0.04 |  |  |
| Male vs Female | 0.965 (0.807, 1.155) | 0.7 | 0.954 (0.798, 1.141) | 0.607 |
| NH Black vs NH White | 1.368 (1.12, 1.672) | 0.002 | 1.326 (1.087, 1.616) | 0.005 |
| Hispanic vs NH White | 1.4 (1.011, 1.939) | 0.043 | 1.326 (0.961, 1.829) | 0.086 |
| Other race/ethnicity vs NH White | 0.899 (0.618, 1.308) | 0.577 | 0.891 (0.612, 1.296) | 0.545 |
| Cancer | 1.085 (0.804, 1.463) | 0.594 | 1.095 (0.812, 1.476) | 0.551 |
| Cardiovascular disease | 1.253 (0.992, 1.583) | 0.059 | 1.248 (0.988, 1.576) | 0.063 |
| Hypertension | 1.357 (1.067, 1.727) | 0.013 | 1.352 (1.064, 1.719) | 0.014 |
| Liver disease | 1.393 (1.021, 1.902) | 0.037 | 1.38 (1.011, 1.884) | 0.042 |
| COPD | 1.869 (1.306, 2.674) | <0.001 | 1.846 (1.291, 2.642) | <0.001 |
| Asthma | 1.3 (1.014, 1.666) | 0.038 | 1.296 (1.011, 1.661) | 0.041 |
| Renal disease | 1.474 (1.092, 1.989) | 0.011 | 1.48 (1.097, 1.997) | 0.01 |
| Diabetes | 1.349 (1.062, 1.713) | 0.014 | 1.34 (1.056, 1.702) | 0.016 |
| Age | 0.991 (0.985, 0.998) | 0.007 | 0.994 (0.988, 1) | 0.042 |

# References

1.   Goldberg Y, Mandel M, Bar-On YM, et al. Waning Immunity after the BNT162b2 Vaccine in Israel. *N Engl J Med*. Dec 9 2021;385(24):e85. doi:10.1056/NEJMoa2114228
2.   Accorsi EK, Britton A, Shang N, et al. Effectiveness of Homologous and Heterologous Covid-19 Boosters against Omicron. *N Engl J Med*. Jun 23 2022;386(25):2433-2435. doi:10.1056/NEJMc2203165
3.   McMurry R, Lenehan P, Awasthi S, et al. Real-time analysis of a mass vaccination effort confirms the safety of FDA-authorized mRNA COVID-19 vaccines. *Med*. Aug 13 2021;2(8):965-978.e5. doi:10.1016/j.medj.2021.06.006
4.   Gianfrancesco MA, Goldstein ND. A narrative review on the validity of electronic health record-based research in epidemiology. *BMC Medical Research Methodology*. 2021/10/27 2021;21(1):234. doi:10.1186/s12874-021-01416-5
5.   Goldstein BA, Phelan M, Pagidipati NJ, Peskoe SB. How and when informative visit processes can bias inference when using electronic health records data for clinical research. *Journal of the American Medical Informatics Association*. 2019;26(12):1609-1617. doi:10.1093/jamia/ocz148
6.   McGee G, Haneuse S, Coull BA, Weisskopf MG, Rotem RS. On the Nature of Informative Presence Bias in Analyses of Electronic Health Records. *Epidemiology*. Jan 1 2022;33(1):105-113. doi:10.1097/ede.0000000000001432
7.   Chodick G, Tene L, Rotem RS, et al. The Effectiveness of the Two-Dose BNT162b2 Vaccine: Analysis of Real-World Data. *Clinical Infectious Diseases*. 2021;74(3):472-478. doi:10.1093/cid/ciab438
8.   Baxter R, Bartlett J, Fireman B, Lewis E, Klein NP. Effectiveness of Vaccination During Pregnancy to Prevent Infant Pertussis. *Pediatrics*. May 2017;139(5)doi:10.1542/peds.2016-4091

9. Lin D-Y, Gu Y, Wheeler B, et al. Effectiveness of Covid-19 Vaccines over a 9-Month Period in North Carolina. *New England Journal of Medicine*. 2022;386(10):933-941. doi:10.1056/NEJMoa2117128
10. Coburn SB, Humes E, Lang R, et al. Analysis of Postvaccination Breakthrough COVID-19 Infections Among Adults With HIV in the United States. *JAMA Network Open*. 2022;5(6):e2215934-e2215934. doi:10.1001/jamanetworkopen.2022.15934
11. Trombetta CM, Kistner O, Montomoli E, Viviani S, Marchi S. Influenza Viruses and Vaccines: The Role of Vaccine Effectiveness Studies for Evaluation of the Benefits of Influenza Vaccines. *Vaccines (Basel)*. May 1 2022;10(5)doi:10.3390/vaccines10050714
12. Teerawattananon Y, Anothaisintawee T, Pheerapanyawaranun C, et al. A systematic review of methodological approaches for evaluating real-world effectiveness of COVID-19 vaccines: Advising resource-constrained settings. *PLoS One*. 2022;17(1):e0261930. doi:10.1371/journal.pone.0261930
13. Korn EL, Graubard BI, Midthune D. Time-to-event analysis of longitudinal follow-up of a survey: choice of the time-scale. *Am J Epidemiol*. Jan 1 1997;145(1):72-80. doi:10.1093/oxfordjournals.aje.a009034
14. Cheung YB, Gao F, Khoo KS. Age at diagnosis and the choice of survival analysis methods in cancer epidemiology. *J Clin Epidemiol*. Jan 2003;56(1):38-43. doi:10.1016/s0895-4356(02)00536-x
15. Thiebaut AC, Benichou J. Choice of time-scale in Cox's model analysis of epidemiologic cohort data: a simulation study. *Stat Med*. Dec 30 2004;23(24):3803-20. doi:10.1002/sim.2098
16. Pencina MJ, Larson MG, D'Agostino RB. Choice of time scale and its effect on significance of predictors in longitudinal studies. *Stat Med*. Mar 15 2007;26(6):1343-59. doi:10.1002/sim.2699